\documentclass[letterpaper, 10 pt, conference]{ieeeconf}  

\usepackage{tikz}
\usepackage{textcomp}
\usepackage{hyperref}
\usepackage{lipsum}

\IEEEoverridecommandlockouts                              

\usepackage{graphics} 
\usepackage{epsfig} 
\usepackage{mathptmx} 
\usepackage{amsmath} 
\usepackage{amssymb}  
\usepackage{float}
\usepackage{multirow}
\usepackage{cite} 

\title{\LARGE \bf
Simulation Framework for Vehicle and Electric Scooter Interaction
}

\author{Zhitong He and Lingxi Li$^*$
\thanks{*Corresponding author.}
\thanks{Zhitong He and Lingxi Li are with the Transportation and Autonomous Systems Institute (TASI), also with the Department of Electrical and Computer Engineering, Indiana University-Purdue University Indianapolis, 723 West Michigan Street, SL-160, Indianapolis, Indiana 46202, USA. Emails:
        {\tt\small \{zh7, LL7\}@iu.edu.}}%
}

\begin{document}
\maketitle

\newcommand\copyrighttext{%
  \footnotesize \textcopyright 2024 IEEE. Personal use of this material is permitted.
  Permission from IEEE must be obtained for all other uses, in any current or future
  media, including reprinting/republishing this material for advertising or promotional
  purposes, creating new collective works, for resale or redistribution to servers or
  lists, or reuse of any copyrighted component of this work in other works.}
\newcommand\copyrightnotice{%
\begin{tikzpicture}[remember picture,overlay]
\node[anchor=south,yshift=10pt] at (current page.south) {\fbox{\parbox{\dimexpr\textwidth-\fboxsep-\fboxrule\relax}{\copyrighttext}}};
\end{tikzpicture}%
}

\copyrightnotice

\thispagestyle{empty}
\pagestyle{empty}

\begin{abstract}

The number of shared micro-mobility services such as electric scooters (e-scooters) has an increasing trend due to the advantages of high efficiency and low cost in short-range travel in urban areas. However, due to the unique characteristics of moving behavior, it is commonly seen that e-scooters may share the road with other motor vehicles. The lack of protection may lead to severe injury for e-scooter riders. The scenario where an e-scooter crosses an intersection or makes a lane change while interacting with an approaching vehicle was commonly seen in real-life traffic data. Such scenarios are hazardous because the intention and behavior of the e-scooter may vary significantly based on the traffic environment conditions. Furthermore, some other vehicles may occlude the presence of the moving e-scooter, which can result in an unexpected collision. In this paper, we propose a simulation platform to mimic the interactions between vehicles and e-scooters. Several traffic scenarios are studied via qualitative and quantitative analysis. The proposed framework is shown to be valuable and efficient for the general risk analysis for vehicle and e-scooter interactions (VEI). 

\end{abstract}

\section{INTRODUCTION}
\label{Sec:Introduction}

\begin{figure}[t]
\centerline{\includegraphics[width=\linewidth]{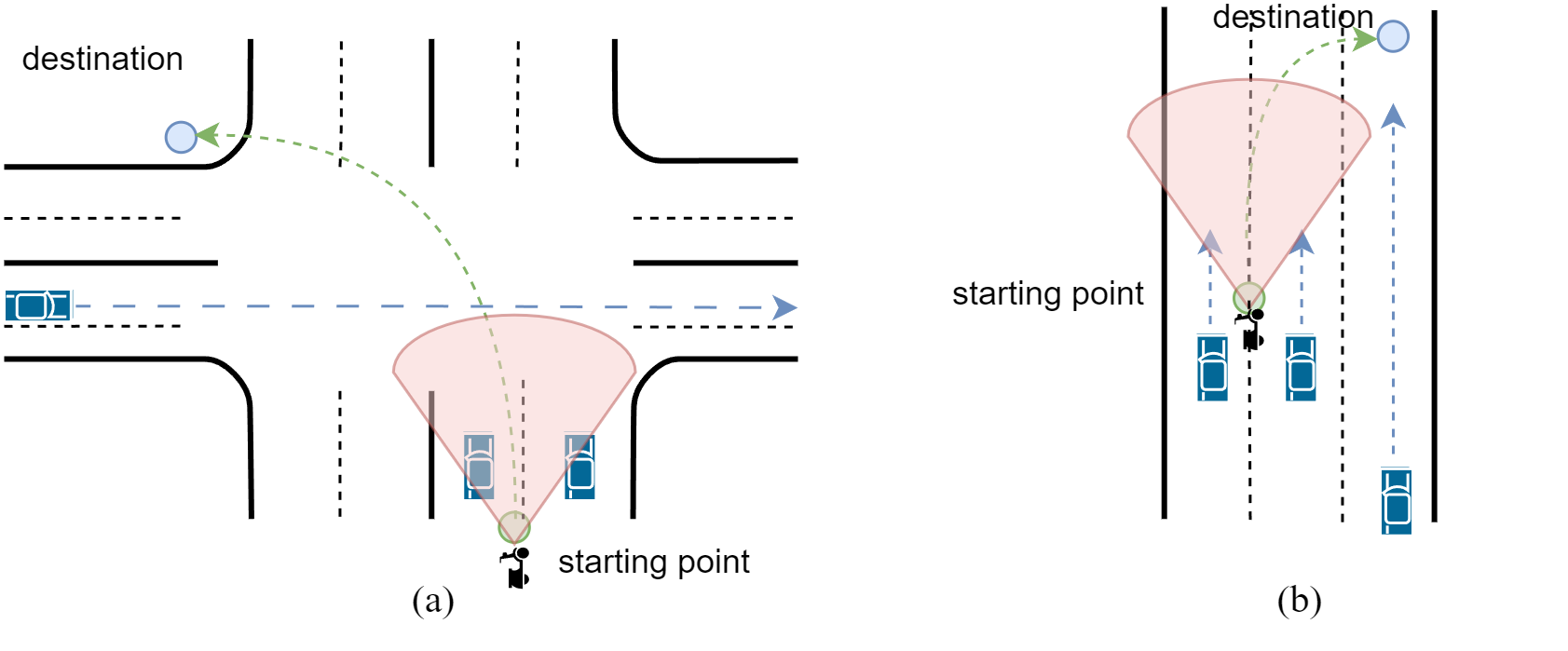}}
\vspace{-2mm}
\caption{Two risky traffic scenarios where the e-scooter wants to cross the intersection with a crossing vehicle is approaching (a) or makes a lane change when a vehicle is approaching from behind (b).}
\label{fig:config}
\end{figure}

The global e-scooter market size was valued at 20.87 million USD in 2021, which is anticipated to continue to grow at a rapid speed \cite{market}. The features of making short trips efficiently and having a comparatively low cost make e-scooters emerge and expand quickly in major cities all over the world. Some existing research investigated the interactions between vehicles and cyclists, which has similar characteristics to e-scooters in some aspects. However, due to the unique moving characteristics that e-scooters may share the road with mobile vehicles and have unpredicted moving intention, it was found that e-scooter crash characteristics do not fully overlap with those of bicycle crashes \cite{shah2021comparison}. The presence of e-scooters could be risky if e-scooter riders do not behave normally under corresponding regulations. The safety research report published by the National Transportation Safety Board (NTSB) indicated an increase in the use of e-scooters and e-bikes, as well as an increase in e-scooter and e-bike rider fatalities and injuries \cite{NTSB}. Shah et al. \cite{shah2021comparison} found that about 10\% of e-scooter-vehicle crashes lead to the injury or fatality of e-scooter riders. Therefore, the interactions between vehicles and e-scooters are critical for traffic safety analysis, which can also be extended to future connected and automated vehicles (CAVs) \cite{Libook}, \cite{teng2023motion},  \cite{JAS}, \cite{chen2021parallel}. Some real-life scenarios also show the potential collision between e-scooters and surrounding vehicles. Several circumstances make the vehicle-e-scooter interactions critical for safety analysis, as demonstrated in Fig.~\ref{fig:config}. The real-life traffic data collected in \cite{prabu2022scendd} also contained these situations. In Fig. 1(a), an e-scooter intends to cross the intersection by passing through two parked vehicles and reaching the destination on the other side. However, a moving vehicle might travel across the intersection at the same time. In Fig.~1(b), an e-scooter plans to make the lane change to reach the destination where a moving vehicle is approaching from behind. Both scenarios are highly risky since the VEI might result in severe consequences.

A method needs to be developed to investigate e-scooter behavior as well as study the interactions between vehicles and e-scooters. The simulation-based experiment is an efficient tool for conducting risk assessment and providing insights into traffic safety improvement. The main contributions of this work are summarized as follows:
\vspace{-2mm}
\begin{enumerate}
    \item A simulation-based framework for studying VEI scenarios is proposed. 
    \item Three risky VEI case studies are conducted using the proposed framework, which includes detailed system design and simulation environment establishment. 
    \item A general risk assessment method is developed for both qualitative and quantitative analysis in order to compare the risk level of different VEI scenarios.
\end{enumerate}

The remainder of this paper is organized as follows: The related works are reviewed in Section~\ref{Sec:Related_works}. Section~\ref{Sec:Framework} introduces the proposed simulation framework. Section~\ref{Sec:Simulation_models} explains the selected simulation models for the vehicle-e-scooter-interaction, including e-scooter and vehicle motion models. Section~\ref{Sec:Simulation_exp} presents the simulation experiments of three use cases and shows qualitative and quantitative simulation results. Finally, the conclusions are drawn in Section~\ref{Sec:Conclusion}.

\section{RELATED WORKS}
\label{Sec:Related_works}

The behavior modeling for e-scooters has been studied using different methodologies. The social force model (SFM) is widely applied in the study of e-scooter behavior as well as research on the interactions between e-scooters and pedestrians. Valero et al. \cite{valero2020adaptation} employed image analysis techniques to extract trajectories of e-scooters and pedestrians from video data. The obtained trajectories serve as the basis for generating scenarios and simulated movements using the SFM. Through a comparison of simulated and experimental trajectories, the study further validated the model and estimated critical parameters using the Cross-Entropy Method (CEM). Similarly, Dias et al. \cite{dias2018simulating} utilized trajectory data gathered from controlled experiments that captured interactions between Segway riders, cyclists, and pedestrians. CEM was also implemented to calibrate the SFM, which is then applied to mixed traffic scenarios under uncongested circumstances. In \cite{liu2022dynamic}, Liu et al. introduced a modified social force model for e-scooters, accounting for essential factors such as kinematic constraints, geometry, and velocity-dependent perception of riders. The comparison of the simulation results between the proposed model and the original model indicated the superior predictive capabilities of the modified SFM, especially for accurately capturing e-scooter interactions with a pedestrian crowd. Except for the model-based e-scooter behavior simulation, Brunner et al. \cite{brunner2020analysis} conducted laboratory experiments by attaching the inertial measurement units (IMUs) to the e-scooters and obtained the movement data with implications for the e-scooter stability feature. The data also captured the kinematic and behavioral characteristics of e-scooter riders.

However, to the best of our knowledge, the interaction between vehicles and e-scooters has not been comprehensively researched. Existing work focuses more on vehicle and pedestrian interaction (VPI) since pedestrians are a typical kind of vulnerable road users in urban traffic. Yang et al. \cite{yang2020multi} designed a multi-state SFM for pedestrian crossing behavior. The proposed model considered the interaction factors influenced by the vehicle and the pedestrian’s individual crossing intention. Simulation experiments demonstrated the feasibility of VPI simulation. In \cite{yang2020social}, Yang et al. also extended the VPI to a more complex scenario, where multiple pedestrians instead of a single pedestrian interact with the moving vehicle. The SFM was further modified based on the variable constraints depending on the effect of the vehicle and the density of surrounding pedestrians. The trajectory data obtained from the controlled experiments were also used to conduct parameter calibration. He et al. \cite{he20235} provided a simulation-based comparison tool for the risk evaluation in the VPI specifically in scenarios where the pedestrian was potentially occluded by some environmental factors. The simulation results verified the effectiveness of the proposed framework for the general risk analysis of VPI. 

Some researchers utilized the partially observable Markov decision process (POMDP) to study VPI. Hsu et al. \cite{hsu2020pomdp} proposed an uncertainty-aware planner for autonomous vehicles to manage the ambiguity of the pedestrian’s intention. The planner regarded the pedestrian intent as a latent variable and the intent is communicated via POMDP to achieve the final decision. Real-life controlled experiments verified the safety and efficiency features of the proposed method. In \cite{lin2019decision}, Lin et al. utilized POMDP to model the traffic situation, which includes making the prediction of other road users. The potential existing vehicles were defined as virtual vehicles at the boundary of areas that exceed the field of view (FOV) and considered in the belief tree search process. A sequence of decisions was obtained from the belief tree search.

The lack of study on VEI is partially due to the fact that most e-scooter riders share the sidewalk with pedestrians or cyclists instead of sharing the road with motor vehicles. Also, e-scooters have a smaller population than pedestrians in the category of vulnerable road users (VRU). Nevertheless, the increasing risk of the potential collision between vehicles and e-scooters is not negligible. Therefore, more research needs to be conducted on improving traffic safety concerning VEI.

\section{Simulation Framework}
\label{Sec:Framework}

\begin{figure}[t]
\centerline{\includegraphics[width=\linewidth]{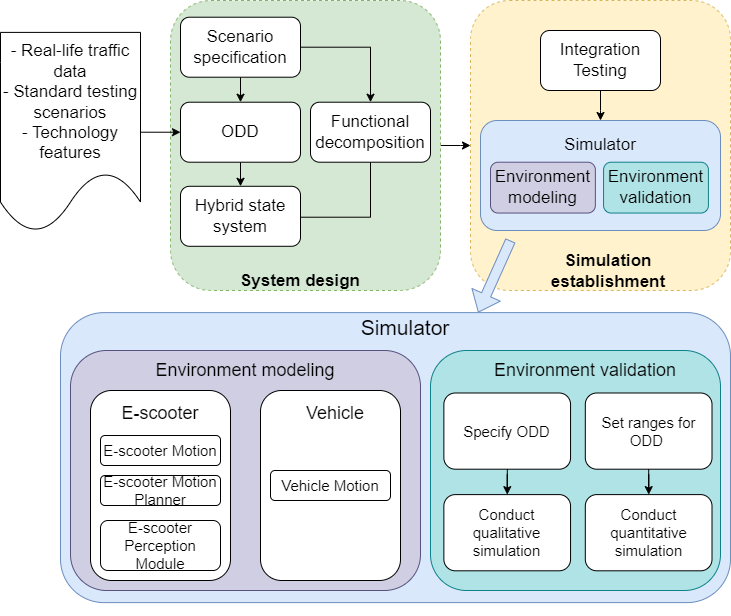}}
\vspace{-2mm}\caption{Framework for the proposed model-based traffic simulation.}
\label{fig:frame}
\end{figure}

The proposed simulation framework contains two major components: system design and simulation establishment, which is shown in Fig.~\ref{fig:frame}. The system design includes the process of defining the operation design domain (ODD) for the VEI process. The simulation establishment demonstrates the detailed simulator structure and the simulation experiments for the proposed system validation.

\vspace{-2mm}
\subsection{System design} 
The designed VEI scenarios are based on the data collection from the previous work \cite{prabu2022scendd}. The scenarios can also be designed by following the protocols of the standard testing scenarios like European New Car Assessment Program (EURO NCAP) VRU testing cases \cite{euroncap}. In this paper, we consider the most commonly seen scenarios where the e-scooter either crosses the intersection or makes a lane change (as illustrated in Fig.~\ref{fig:config}). The ODD design for these two circumstances varies slightly due to the difference in the behavior of the interacting agents.

\subsection{Simulation establishment}

The simulation establishment stage contains the detailed simulation model setup. This paper concentrates on the e-scooter moving behavior and relevant modules such as e-scooter perception and motion. Both the e-scooter and vehicle models are appropriately designed to achieve realistic simulation and save computational power. Integration testing is implemented to test the functionality of cooperation between different modules. The detailed formulation and parameters setup will be introduced in the next section. After the modeling step, qualitative and quantitative simulations are conducted to validate the system’s effectiveness. Additionally, the simulation results can depict some risk factors in the target VEI scenarios.




\section{Simulation Models}
\label{Sec:Simulation_models}


To simulate the interaction between e-scooters and vehicles, we start by selecting the appropriate motion models for them. An e-scooter behavior model is established based on a state transition model and a simplified social force model. We also use a generic perception model to represent the sensing module of the e-scooter. 
\vspace{-2mm}
\subsection{E-scooter Model}

\subsubsection{E-scooter Dynamics}

The social force model \cite{yang2020multi} is used to apply the longitudinal and lateral forces on a point-mass Newtonian dynamic.
\vspace{-2mm}
\begin{equation}
\ddot{X_{esc}}=\frac{1}{m_{esc}} f_{total},\label{eq:pointmass}
\end{equation}

\noindent where ${X_{esc}}\in \mathbb{R}^{4}, X_{esc}:=[s_{esc_x}, s_{esc_y}, v_x, v_y]^\top$ is the e-scooter state vector containing the position and velocity along the $x$ and $y$ axis, respectively, $m_{esc}$ is the e-scooter mass, and $f_{total}$ is the total applied force. 

\subsubsection{Social Force Crossing/Lane-changing Model}

For the e-scooter movement model, we integrate a simplified applied social force model \cite{valero2020adaptation} into a pre-defined finite state machine (FSM). The applied force is divided into two categories: propelling force and repulsive force. The accumulated influence of the interaction is the
summation of all individual forces. In the model, the total force was designed as follows:
\vspace{-2mm}
\begin{equation}
f_{total}=f_{des} + \sum {f_{veh}},\label{eq:ftotal}
\end{equation}

\noindent  where $f_{des}$ is the propelling force that can be regarded as a driving force to motivate the e-scooter to approach the destination. In contrast, $\sum  f_{veh}$ is the summation of the repulsive force provided by each vehicle functioning as a resistance force that prevents the e-scooter from moving toward the influential object.

The destination force $f_{des}\in \mathbb{R}^{2}$ can be defined as
\vspace{-2mm}
\begin{equation}
{f_{des}}={k_{des}}({v_{des}} - v_{esc}),\label{eq:fdes}
\end{equation}

\noindent where $k_{des}$ is a parameter that scales the difference between $v_{des}$ and $v_{esc}$.  $v_p\in \mathbb{R}^{2}$ is the current e-scooter velocity vector and  $v_d\in \mathbb{R}^{2}$ is the desired velocity vector, defined by
\vspace{-2mm}
\begin{equation}
v_d:=v_0 \frac{s_{des}-s_{esc}}{\left|s_{des}-s_{esc}\right|^{2}+\sigma_{des}^2}\label{eq:desv},
\end{equation}

\noindent where $s_{des}\in \mathbb{R}^{2}$ is the destination coordinates vector, $s_{esc}\in \mathbb{R}^{2}$ is the e-scooter current position, $\sigma_{des}$ is a scalar that adjusts $v_d$ depending on the e-scooter's distance to the destination. The speed $v_0$ is a constant parameter that represents the most commonly seen velocity for e-scooters.

The effective virtual vehicle force $f_{veh}\in \mathbb{R}^{2}$ is defined as:
\vspace{-2mm}
\begin{equation}
{\sum f_{veh}}=\sum_i {A_{veh}}\cdot exp(-b_{veh}.d_{v_i2esc})\cdot \Vec{n}_{v_i2esc},\label{eq:fveh}
\end{equation}

\noindent where $A_{veh}$ and $b_{veh}$ are the pre-defined parameters used for the calculation. $d_{v_i2esc}$ is the distance between the $i$-th vehicle's influential point $s_{i,influence}$ and the current e-scooter position $s_{esc}$. $\Vec{n}_{v_i2esc}$ is a unit vector showing the direction from $s_{i,influence}$ to $s_{esc}$. The summation of each individual vehicle force is the total repulsive force that applied on the e-scooter.

\begin{figure}[t]
\centerline{\includegraphics[scale=0.13]{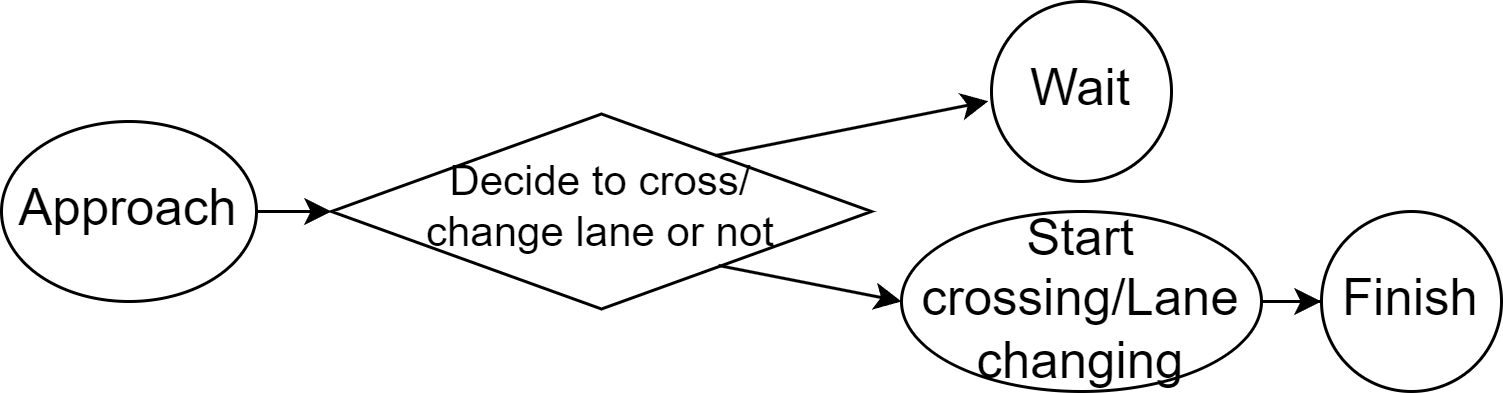}}
\caption{State transition of the motion planner for the e-scooter.}
\label{fig:fsm}
\end{figure}

\subsubsection{Perception Model}

We use a simple geometric model to represent the perception module. We use a sector to represent the field of view (FOV) of the e-scooter in a 2D plane. When a vehicle intersects with the sector, it indicates that it is within the FOV's range. Thus, the perceived vehicle can be regarded as an influential vehicle to provide the propelling force that keeps the e-scooter away. In the example scenarios shown in Fig.~\ref{fig:config}, the two parked cars in the intersection scenario can be seen by the e-scooter. In contrast, no car can be detected by the FOV in the straight road scenario. The occlusions are not considered in the simulation for computational simplicity.

\subsubsection{State Transition Model} \label{fsm}

The proposed FSM shown in Fig.~\ref{fig:fsm} illustrates the state transition when an e-scooter would like to cross the intersection or make a lane change. The e-scooter first approaches a position to make the decision of crossing or changing lanes or staying still. The decision is made depending on the e-scooter's moving threshold. For example, in the intersection crossing scenario shown in Fig.~\ref{fig:config}a, if the gap distance is smaller than the crossing threshold, the e-scooter will stay in the decision-making position. Otherwise, the e-scooter will decide to move forward and start the crossing or lane changing until it reaches the destination. 

\vspace{-2mm}

\subsection{Vehicle Motion model}

The motion of the vehicle in this paper is represented by a kinematic bicycle model. The vehicle state vector $X_{veh}=[x_{veh}, y_{veh}, \psi, v_{veh}]^\top$ includes four individual states of longitudinal position, lateral position, heading angle, and velocity, respectively.  Details of the kinematic bicycle model can be found in \cite{kong2015kinematic}. The current vehicle model in the example VEI system does not consider lateral control or braking behavior. Therefore, the moving vehicles are designed to follow the pre-defined trajectories at a constant speed in all simulations.



\section{Simulation Experiments}
\label{Sec:Simulation_exp}

\begin{figure}
\centerline{\includegraphics[scale=0.4]{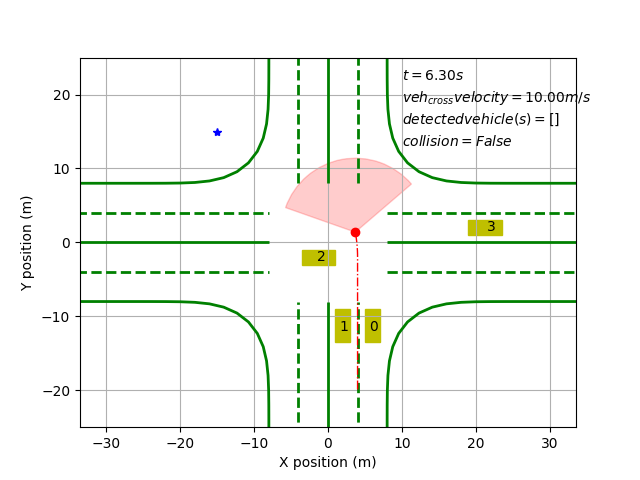}}
\vspace{-2mm}\caption{Illustration of the vehicle e-scooter interaction scenario in an intersection with two crossing vehicles. The red dot shows the position of the e-scooter. The red sector represents the FOV of the e-scooter. The blue star is the destination that the e-scooter targets to reach. The yellow boxes represent individual vehicles.}
\label{fig:twoveh}
\end{figure}

One scenario of a four-lane intersection and one of a four-lane straight road were created for the simulation experiments with single-lane width of 4 m. In the intersection scenario, the initial position of the e-scooter was selected to be behind the two parked vehicles. The destination was set to be at the corner on the opposite side of the intersection. The e-scooter will make the decision of crossing or not based on the FSM introduced in Section \ref{fsm}. A vehicle was moving from the West to the East from the bird’s eye view. The crossing vehicle followed a pre-defined straight trajectory at a constant speed and may or may not encounter the e-scooter. A variant of the intersection scenario would include a second crossing vehicle traveling from the East to the West following a straight trajectory with a fixed velocity as presented in Fig.~\ref{fig:twoveh}. The existence of both crossing vehicles made the interaction more dangerous since the virtual e-scooter could collide with either vehicle.

In the straight road scenario, the e-scooter started from the initial position, which is in front of the two parallel moving vehicles, and would make the decision to change lanes based on the crossing threshold. The two vehicles behind moved at a constant low speed and would not catch up with the e-scooter. A third moving vehicle was marching at a higher speed in the adjacent lane. The destination of the e-scooter was placed to the North-East of the initial position.

\begin{table}[]
\caption{Variables for Qualitative Simulations}
\begin{center}
\begin{tabular}{|cccc|}
\hline
\multirow{2}{*}{Symbol} & Intersection & Straight Road& \multirow{2}{*}{Unit} \\ \cline{2-3}
                        & Value            & Value           &                       \\ \hline
$(x_{veh_0},\hspace{2pt} y_{veh_0})$       &(6.0,\hspace{2pt}-12.0)&(-6.0,\hspace{2pt}-24.0)&  $m$   \\
$(x_{veh_1},\hspace{2pt} y_{veh_1})$       &(2.0,\hspace{2pt}-16.0)&(-2.0,\hspace{2pt}-24.0)&  $m$     \\
$(x_{veh_2,init},\hspace{2pt}y_{veh_2,init})$  &(-75.0,\hspace{2pt}-2.0)&(2.0,\hspace{2pt}-60.0)&  $m$  \\
$^{*}(x_{veh_3,init},\hspace{2pt}y_{veh_3,init})$ &(85.0,\hspace{2pt}2.0)&    /    &   $m$     \\
$v_{veh_0}$                 &      /            &   1.0     &   $m/s$      \\
$v_{veh_1}$                 &      /            &    1.0      &  $m/s$         \\
$v_{veh_2}$                 &  10.0   &       10.0     &     $m/s$     \\
$^{*}v_{veh_3}$             &  10.0   &       /     &         $m/s$         \\ \hline
$(x_{esc,init},\hspace{2pt}y_{esc,init})$  &(4.0,\hspace{2pt}-30.0)&   (-4.0,\hspace{2pt}-10.0)&   $m$                  \\
$(x_{des},\hspace{2pt}y_{des})$            &(-15.0,\hspace{2pt}15.0)&  (2.0,\hspace{2pt}20.0)  &    $m$                \\
$r_{esc,FOV}$           &    10.0      &    10.0     &      $m$             \\
$\alpha_{esc,FOV}$      &    120       &    120      &      $deg$              \\ \hline
\end{tabular}%
\label{tab:par_qual}
\end{center}
\end{table}

\vspace{-2mm}
\subsection{Qualitative Analysis}

\begin{figure*}[t!]
\centerline{\includegraphics[width=0.95\linewidth]{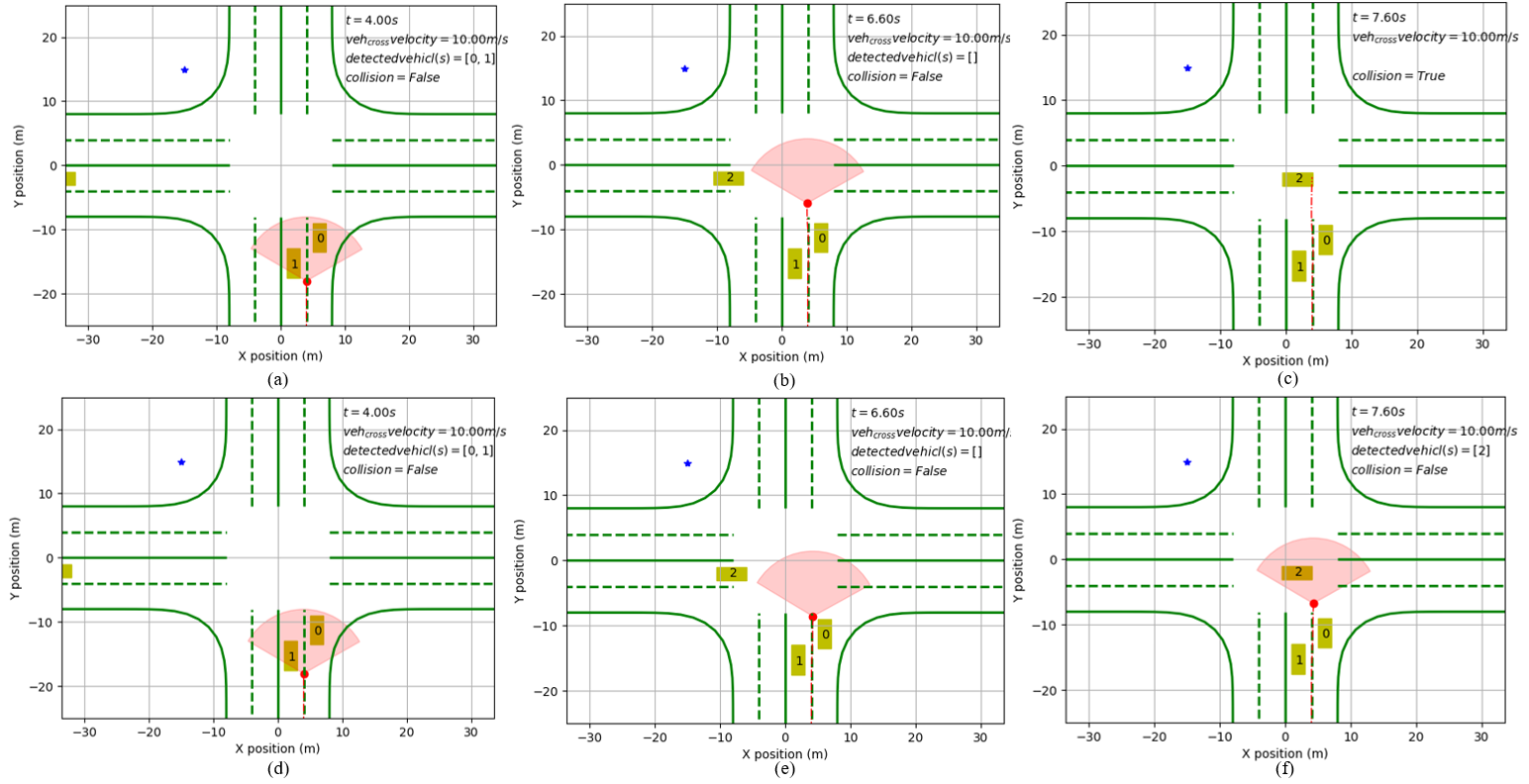}}
\vspace{-2mm}\caption{Screenshots of the simulation in the intersection scenario with {\it Aggressive} and {\it Normal} types of e-scooters. The interaction process of an {\it Aggressive} e-scooter is shown in the first row of the graph. The e-scooter first approached the waiting position and started to cross the interaction while the moving vehicle began crossing the intersection at a much higher speed. In the end, the collision occurred at $t=7.6s$. The second row of the graph shows the interaction process of a {\it Normal} e-scooter. Due to the higher crossing threshold, the e-scooter kept waiting at the decision-making position. The e-scooter kept a safe distance from the crossing vehicle and avoided a potential collision.}

\label{fig:qual_results_vehcross}
\end{figure*}

\begin{figure*} 
\centerline{\includegraphics[width=0.95\linewidth]{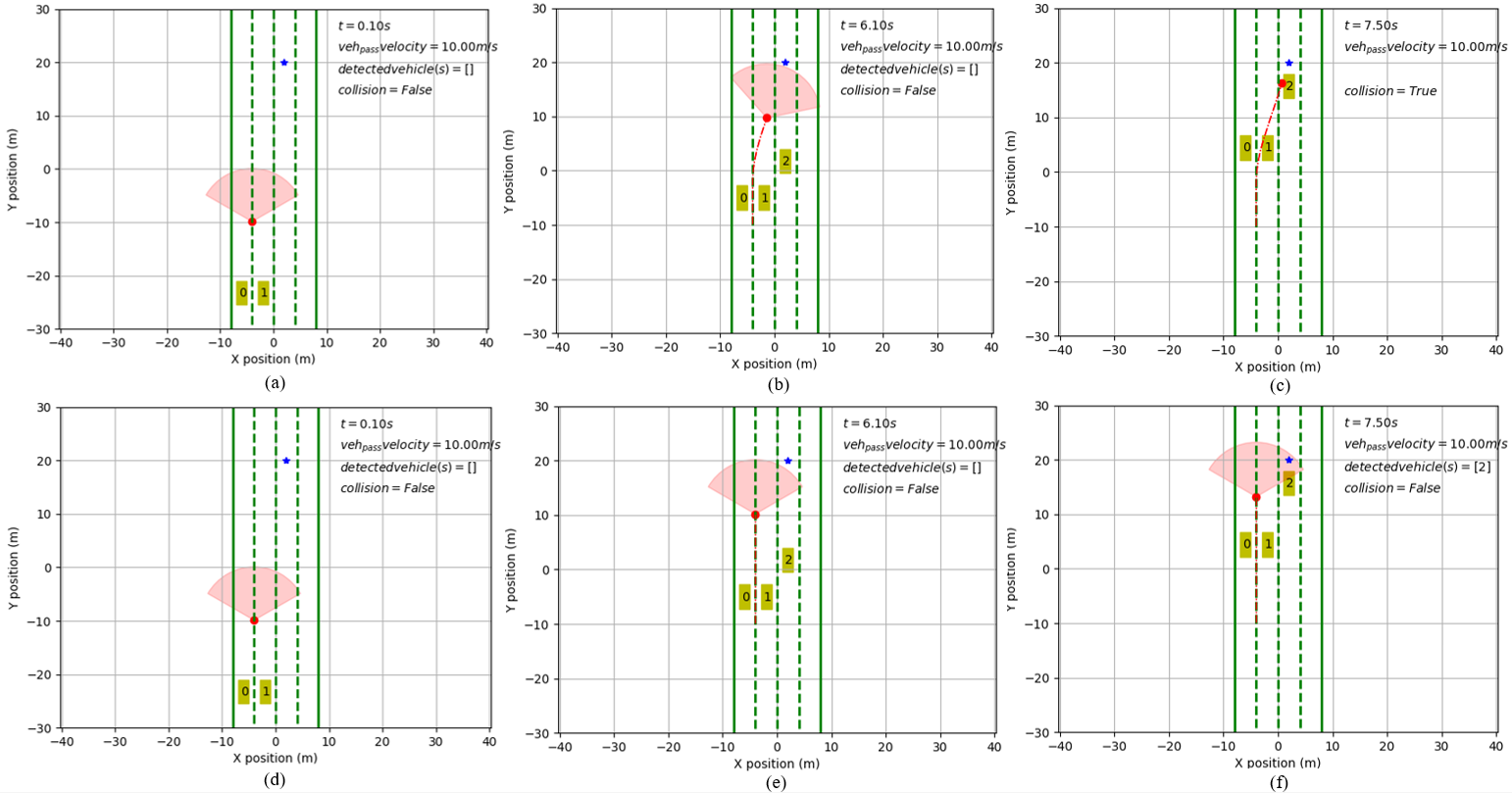}}
\vspace{-2mm}\caption{Screenshots of the simulation in the straight road scenario with {\it Aggressive} and {\it Normal} types of e-scooters. In the first row, the {\it Aggressive} e-scooter started changing lanes before $t=6.1s$ and had a crash with the passing vehicle coming from behind at $t=7.5s$. The simulation sequence of the second row illustrates a {\it Normal} e-scooter kept waiting in the current lane and did not collide with the passing vehicle.}

\label{fig:qual_results_vehpass}
\end{figure*}

The simulation experiments started with three sets of control experiments to test the results of the crossing and lane-changing processes of different types of e-scooters. We define the e-scooter type as {\it Aggressive} and {\it Normal}. The {\it Aggressive} one has more willing to cross or make the lane change, while the {\it Normal} one only moves when the gap is comparatively large. The parameters for the intersection and straight road scenarios are listed in Table~\ref{tab:par_qual}. The parameters with a $*$ were configured for the two-vehicle crossing intersection. The velocity for all crossing and approaching vehicles is 10 $m/s$. In the pair tests of each scenario, all the variables are set the same. The only difference is the e-scooter type. For simplicity, we configure that the {\it Aggressive} type uses a low crossing threshold and {\it Normal} type uses a high crossing threshold.

Screenshots of two episodes of the intersection scenario with one crossing vehicle are illustrated in Fig.~\ref{fig:qual_results_vehcross}. The crossing gap was larger than the threshold of the {\it Aggressive} type of e-scooter, and the e-scooter decided to cross the intersection but collided with the crossing vehicle at the end of the simulation. While for the {\it Normal} type, the e-scooter approached the decision-making position the same as the {\it Aggressive} one but did not cross the intersection and stayed still due to the high crossing threshold value. Concerning the two-vehicle crossing scenario, it has a similar result that the {\it Aggressive} e-scooter collided with the crossing vehicle while no collision happened to the {\it Normal} e-scooter. The two episodes' simulation of the {\it Aggressive} and {\it Normal} e-scooters in the lane changing scenario (shown in Fig.~\ref{fig:qual_results_vehpass}) depicts a similar interaction consequence. The {\it Aggressive} e-scooter did not know the existence of the passing vehicle and collided with it after changing the lane. However, for the {\it Normal} e-scooter, because of the more conservative behavior, the e-scooter stayed in the current lane and successfully avoided the potential crash.

\vspace{-2mm}
\subsection{Quantitative Analysis}

\begin{figure}[H]
\centerline{\includegraphics[width=1.0\linewidth]{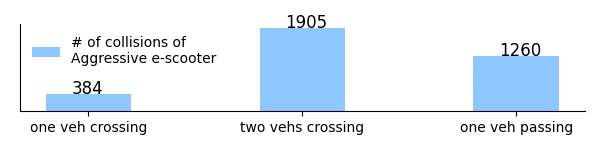}}
\vspace{-4mm}
\caption{Simulation results for e-scooters in three use cases.}
\label{fig:histogram}
\end{figure}

The quantitative simulation aims to provide more verification of the VEI simulation. We conducted experiments for each scenario with {\it Normal} type and {\it Aggressive} type of e-scooter, respectively. The variables of the experiments for three scenarios are shown in Table~\ref{tab:par_quant}. We can see from Fig.~\ref{fig:histogram} and Table~\ref{tab:col_quant} that the two-vehicle crossing case had the largest number of collisions as well as the highest collision rate. In addition, no collision occurred when the e-scooter type was {\it Normal} in all three scenarios. This is because the crossing threshold was defined as a static value. If the e-scooter arrived at the decision-making position and decided not to cross the intersection or not to make the lane change, then the e-scooter would keep waiting until the end of the simulation episode. The parked vehicles in the intersection scenario or the following vehicles in the straight road scenario would not collide with the e-scooter by the system design.

\begin{table}[]
\caption{Variables for Quantitative Simulations}
\begin{center}
\begin{tabular}{|ccccc|}
\hline
\multirow{2}{*}{Symbol} & Intersection & Straight road& \multirow{2}{*}{Step size} & \multirow{2}{*}{Units} \\ \cline{2-3}
                        & Value            & Value           &                            &                        \\ \cline{1-5}
$y_{veh_0}$    & [-16.0,\hspace{2pt}-11.0]         &   /  &      5.0     &$m$                  \\
$y_{veh_1}$     & [-16.0,\hspace{2pt}-11.0]        &   /     & 5.0       &$m$                   \\
$y_{veh_2,init}$  &   /     & [-75.0,\hspace{2pt}-55.0]       & 5.0      &$m$                  \\
$x_{veh_2,init}$   & [-85.0,\hspace{2pt}-65.0]  &   / & 10.0     &$m$                      \\
$^{*}x_{veh_3,init}$ & [75.0,\hspace{2pt}95.0]   &   /    & 10.0        & $m$              \\
$v_{veh_2}$          & [10.0,\hspace{2pt}15.0] & [10.0,\hspace{2pt}15.0]& 5.0/2.5     & $m/s$    \\
$^{*}v_{veh_3}$      & [10.0,\hspace{2pt}15.0]  &   /   & 5.0    & $m/s$         \\ \hline
$y_{esc,init}$       & [-30.0,\hspace{2pt}-20.0] & [-15.0,\hspace{2pt}-10.0]  & 5.0/2.5  & $m$   \\
$x_{des}$  & [-15.0,\hspace{2pt}-10.0] & [4.0,\hspace{2pt}9.0] & 5.0/1.0      & $m$           \\
$y_{des}$   & [12.0,\hspace{2pt}17.0] & [17.0,\hspace{2pt}22.0]   & 5.0/1.0   & $m$        \\
$r_{esc,FOV}$    & [10.0,\hspace{2pt}20.0]& [10.0,\hspace{2pt}20.0] & 5.0         & $m$    \\
$\alpha_{esc,FOV}$ & [60,\hspace{2pt}120]& [60,\hspace{2pt}120]& 30       & $deg$            \\ \hline
\end{tabular}
\label{tab:par_quant}
\end{center}
\end{table}

\vspace{-4mm}
\begin{table}[]
\caption{Collision Rates in Quantitative Simulations}
\resizebox{\linewidth}{!}{%
\begin{tabular}{cccc}
\begin{tabular}[c]{@{}c@{}}Use case \end{tabular} & \begin{tabular}[c]{@{}c@{}}Collision rate of\\ {\it Aggressive} type of e-scooter (\%)\end{tabular} & \begin{tabular}[c]{@{}c@{}}Collision rate of \\{\it Normal} type of e-scooter (\%)\end{tabular}\\ \hline
One vehicle crossing scenario    & 22.22      & 0.00         \\ 
Two vehicles crossing scenario   & 36.75      & 0.00          \\ 
One vehicle passing scenario     & 18.67      & 0.00                                                                                                                                               
\end{tabular}\label{tab:col_quant}
}
\end{table}

Furthermore, the collision rates shown in Table~\ref{tab:col_quant} partially indicate the risk level of different scenarios. Compared with the original scenario, one more crossing vehicle was added in the two-vehicle crossing scenario, where the rest configurations were kept the same.  The two-vehicle crossing use case has an increased collision rate of 14.53 \%, which is reasonable because the e-scooter may collide with vehicles traveling from both East and West directions. 
The collision rate in the e-scooter lane-changing scenario is the lowest at 18.67\%. But it does not necessarily reveal the scenario is safer than the other two. Because the interaction between the vehicle and the e-scooter for different scenarios mainly depends on the system design and configuration.

\vspace{-1mm}
\section{Conclusion}
\label{Sec:Conclusion}

This paper introduced a general simulation framework for vehicle-e-scooter interactions. In the proposed framework, a hybrid-state traffic system was designed based on real-life traffic data. The simulation was established by setting up environmental models, including vehicle and e-scooter models. Each road user contained an individual motion module and motion planner. After defining the ODD for the three use cases, qualitative and quantitative analyses were implemented to verify the feasibility of using the model-based method for the target traffic simulation. The simulation results depicted the risk level of different vehicle-e-scooter-interaction scenarios. The potential risk factors can be drawn by the simulation.
Generally, the simulation framework is proven to be valid when studying traffic scenarios where an e-scooter encounters multiple static or dynamic vehicles. 

For future work, the collision avoidance planner of the moving vehicle with crash probability needs to be developed for a more realistic VEI simulation. In addition, more road users, such as pedestrians and mobility vehicles of different sizes, may be considered to create a more complex traffic environment for further vehicle and VRU interaction research.

\addtolength{\textheight}{-12cm}   



\bibliography{ref}

\begin{thebibliography}{10}

\bibitem{market}
P.~M. Research, ``{Electric Scooter Market Share, Size, Trends, Industry Analysis Report}.'' \url{https://www.polarismarketresearch.com/industry-analysis/electric-scooter-market}, 2021.
\newblock [Online].

\bibitem{shah2021comparison}
N.~R. Shah, S.~Aryal, Y.~Wen, and C.~R. Cherry, ``Comparison of motor vehicle-involved e-scooter and bicycle crashes using standardized crash typology,'' {\em Journal of safety research}, vol.~77, pp.~217--228, 2021.

\bibitem{NTSB}
N.~T.~S. Board, ``{Micromobility: Data Challenges Associated with Assessing the Prevalence and Risk of Electric Scooter and Electric Bicycle Fatalities and Injuries}.'' \url{https://www.ntsb.gov/safety/safety-studies/Documents/SRR2201.pdf}, 2022.
\newblock [Online].

\bibitem{Libook}
Y.~Chen and L.~Li, {\em Advances in Intelligent Vehicles}.
\newblock Academic Press, 2013.

\bibitem{teng2023motion}
S.~Teng, X.~Hu, P.~Deng, B.~Li, Y.~Li, Y.~Ai, D.~Yang, L.~Li, Z.~Xuanyuan, F.~Zhu, {\em et~al.}, ``Motion planning for autonomous driving: The state of the art and future perspectives,'' {\em IEEE Transactions on Intelligent Vehicles}, 2023.

\bibitem{JAS}
J.~Tan, C.~Xu, L.~Li, F.-Y. Wang, D.~Cao, and L.~Li, ``Guidance control for parallel parking tasks,'' {\em IEEE/CAA Journal of Automatica Sinica}, vol.~7, no.~1, pp.~301--306, 2020.

\bibitem{chen2021parallel}
L.~Chen, X.~Hu, G.~Wang, D.~Cao, L.~Li, and F.-Y. Wang, ``Parallel mining operating systems: From digital twins to mining intelligence,'' in {\em 2021 IEEE 1st International Conference on Digital Twins and Parallel Intelligence (DTPI)}, pp.~469--473, IEEE, 2021.

\bibitem{prabu2022scendd}
A.~Prabu, N.~Ranjan, L.~Li, R.~Tian, S.~Chien, Y.~Chen, and R.~Sherony, ``Scendd: A scenario-based naturalistic driving dataset,'' in {\em 2022 IEEE 25th International Conference on Intelligent Transportation Systems (ITSC)}, pp.~4363--4368, IEEE, 2022.

\bibitem{valero2020adaptation}
Y.~Valero, A.~Antonelli, Z.~Christoforou, N.~Farhi, B.~Kabalan, C.~Gioldasis, and N.~Foissaud, ``Adaptation and calibration of a social force based model to study interactions between electric scooters and pedestrians,'' in {\em 2020 IEEE 23rd International Conference on Intelligent Transportation Systems (ITSC)}, pp.~1--7, IEEE, 2020.

\bibitem{dias2018simulating}
C.~Dias, H.~Nishiuchi, S.~Hyoudo, and T.~Todoroki, ``Simulating interactions between pedestrians, segway riders and cyclists in shared spaces using social force model,'' {\em Transportation research procedia}, vol.~34, pp.~91--98, 2018.

\bibitem{liu2022dynamic}
Y.-C. Liu, A.~Jafari, J.~K. Shim, and D.~A. Paley, ``Dynamic modeling and simulation of electric scooter interactions with a pedestrian crowd using a social force model,'' {\em IEEE Transactions on Intelligent Transportation Systems}, vol.~23, no.~9, pp.~16448--16461, 2022.

\bibitem{brunner2020analysis}
P.~Brunner, A.~L{\"o}cken, F.~Denk, R.~Kates, and W.~Huber, ``Analysis of experimental data on dynamics and behavior of e-scooter riders and applications to the impact of automated driving functions on urban road safety,'' in {\em 2020 IEEE Intelligent Vehicles Symposium (IV)}, pp.~219--225, IEEE, 2020.

\bibitem{yang2020multi}
D.~Yang, K.~Redmill, and {\"U}.~{\"O}zg{\"u}ner, ``A multi-state social force based framework for vehicle-pedestrian interaction in uncontrolled pedestrian crossing scenarios,'' in {\em 2020 IEEE Intelligent Vehicles Symposium (IV)}, pp.~1807--1812, IEEE, 2020.

\bibitem{yang2020social}
D.~Yang, {\"U}.~{\"O}zg{\"u}ner, and K.~Redmill, ``A social force based pedestrian motion model considering multi-pedestrian interaction with a vehicle,'' {\em ACM Transactions on Spatial Algorithms and Systems (TSAS)}, vol.~6, no.~2, pp.~1--27, 2020.

\bibitem{he20235}
Z.~He, L.~Capito, K.~Redmill, F.~{\"O}zg{\"u}ner, and {\"U}.~{\"O}zg{\"u}ner, ``5 risk analysis for vehicle--pedestrian interaction with extended sensing,'' {\em Towards Human-Vehicle Harmonization}, vol.~3, p.~65, 2023.

\bibitem{hsu2020pomdp}
Y.-C. Hsu, S.~Gopalswamy, S.~Saripalli, and D.~A. Shell, ``A pomdp treatment of vehicle-pedestrian interaction: Implicit coordination via uncertainty-aware planning,'' in {\em 2020 IEEE/RSJ International Conference on Intelligent Robots and Systems (IROS)}, pp.~1984--1991, IEEE, 2020.

\bibitem{lin2019decision}
X.~Lin, J.~Zhang, J.~Shang, Y.~Wang, H.~Yu, and X.~Zhang, ``Decision making through occluded intersections for autonomous driving,'' in {\em 2019 IEEE Intelligent Transportation Systems Conference (ITSC)}, pp.~2449--2455, IEEE, 2019.

\bibitem{euroncap}
EURONCAP, ``{AEB/LSS VRU Systems Test protocol v4.3}.'' \url{https://cdn.euroncap.com/media/75436/euro-ncap-aeb-lss-vru-test-protocol-v43.pdf}, 2022.
\newblock [Online].

\bibitem{kong2015kinematic}
J.~Kong, M.~Pfeiffer, G.~Schildbach, and F.~Borrelli, ``Kinematic and dynamic vehicle models for autonomous driving control design,'' in {\em 2015 IEEE intelligent vehicles symposium (IV)}, pp.~1094--1099, IEEE, 2015.

\end{thebibliography}

\end{document}